\documentclass[10pt]{article}
\usepackage{amsfonts}
\usepackage{amsmath}
\usepackage{amssymb}

\begin{document}

\title{Conformal invariance of massless Duffin-Kemmer-Petiau theory \\
in Riemannian space-times}
\author{R. Casana$^{a}$\thanks{%
casana@ift.unesp.br}, J.T. Lunardi$^{b}$\thanks{%
jttlunardi@uepg.br}, B. M. Pimentel$^{a}$\thanks{%
pimentel@ift.unesp.br} and R. G. Teixeira$^{c}$\thanks{%
randall@cce.ufes.br}}
\date{}
\maketitle

\vspace{-1cm}

\begin{center}
{\small \textit{$^a$ Instituto de F\'{\i}sica Te\'orica,
Universidade
Estadual Paulista}}\\[0pt]
{\small \textit{Rua Pamplona 145, CEP 01405-900, S\~ao Paulo, SP,
Brazil}}

\vspace{.2cm}

{\small \textit{$^b$ Grupo de F\'{\i}sica Te\'orica. Departamento
de
Matem\'atica e Estat\'{\i}stica}}\\[0pt]
{\small \textit{Universidade Estadual de Ponta Grossa. Av. Gal.
Carlos
Cavalcanti 4748}}\\[0pt]
{\small \textit{84032-900, Ponta Grossa, PR, Brazil}}

\vspace{.2cm}

{\small \textit{$^c$ Departamento de F\'{\i}sica, Universidade Federal do Esp%
\'{\i}rito Santo }}\\[0pt]
{\small \textit{Av. Fernando Ferrari s/n, Goiabeiras, CEP 29060-900, Vit\'{o}%
ria, ES, Brazil}}

\end{center}

\begin{abstract}
We investigate the conformal invariance of massless
Duffin-Kemmer-Petiau theory coupled to riemannian space-times. We
show that, as usual, in the minimal coupling procedure only the spin
1 sector of the theory -which corresponds to the electromagnetic
field- is conformally invariant. We   {show also} that the conformal
invariance of the spin 0 sector can be naturally achieved by
introducing a compensating term in the lagrangian. Such a procedure
-besides not modifying the spin 1 sector- leads to the well-known
conformal coupling between  {the} scalar curvature and the massless
Klein-Gordon-Fock field. Going beyond the riemannian spacetimes, we
 {briefly discuss} the effects of a nonvanishing torsion in
the scalar case.
\end{abstract}


\section{Introduction}


The conformal symmetry was first introduced into physics in the beginning of
the last century when Cunningham and Bateman \cite{cunningham} showed that
Maxwell's vacuum equations are also invariant under the larger conformal
group, which proves to be the maximal extension of the Poincar\'e group that
leaves the light-cone invariant. A particular feature of the conformal group
is that it extends the usual Lorentz group in allowing transformations to
frames undergoing constant acceleration. Some aspects of conformal
invariance on curved space-time were studied in \cite{fulton}.

The conformal symmetry of a quantum field theory was intensively
studied at the end of sixties and the seventies, with the motivation
coming independently from the Bjorken scaling hypothesis \cite{Bj}
and the theory of second order phase transitions \cite{lista1}. It
turns out that conformal symmetry of a two-dimensional quantum field
theory implies the existence of infinitely many conservation laws
and strong constraints on the correlation functions. In particular,
spectacular results were obtained in statistical mechanical models
in two dimensions \cite{lista2}. The subject of two-dimensional
conformal field theories (CFT) has become very fashionable due to
its connections with string theory and quantum gravity
\cite{Polyakov,buchbinder}, and {it} has grown into a separate
branch of mathematical physics with its own methods and own language
\cite{ketov}. In some 4-dimensional theories with supersymmetry the
$\beta$-function vanishes (to lowest order or to all orders) and
some features reminiscent of two-dimensional CFT emerge. This
subject attracted a lot of interest, enhanced in recent years in
connection with the conjecture about AdS/CFT correspondence (see
\cite{maldacena} for a review).

The Duffin-Kemmer-Petiau (DKP) theory \cite{dkp} provides a unified
description of scalar and vector fields through a first order
equation, similar to the Dirac's one. Notwithstanding the complete
equivalence between the alternative treatments based on DKP and
Klein-Gordon-Fock (KGF) and Proca equations in the free field case,
such equivalence does not generally holds when interactions are
taken into account \cite{refdkp}. For instance, the
massive DKP theory is richer than KGF and Proca theories when \textit{%
minimally coupled} \cite{misner,hehl} to a Riemann--Cartan spacetime \cite%
{spin0,spin1}. On the other hand, the Harish-Chandra theory for massless DKP
fields in Minkowski spacetime \cite{HarishC} was extended in references \cite%
{ijmpa,CQG} to curved spacetimes via minimal coupling procedure and
it was proved the complete equivalence  {between this theory and the
KGF and Maxwell ones}, both in  {riemannian} and in Riemann-Cartan
space-times.

Our aim in this work is  {the study of the properties of the
massless DKP field in riemannian space-times under conformal
transformations}. We start from actions which are invariant under
general coordinate transformations and, as it is well known, in this
case the conformal invariance of the theory can be inferred from its
invariance under the so-called Weyl rescalings
\cite{buchbinder,blagojevic}. The paper is organized as follows. In
section 2 we briefly review the massless DKP theory minimally
coupled to riemannian space-times, quoting the main results we
{shall} need. In section 3 we set the transformation laws {for} DKP
fields under Weyl transformations and show that only the spin 1
sector of this theory is conformally invariant. We achieve the
conformal invariance of also the spin 0 sector through the
introduction of a compensating term -{which by its turn} does not
modify the spin 1 sector- in the lagrangian, {thus} recovering in a
natural way the usual (non-minimal) conformal coupling between the
scalar massless KGF field and the scalar curvature. At the end of
this section we go beyond the riemannian {space-times} and briefly
discuss the effects of a nonvanishing torsion on the scalar sector.
Our conclusions are presented in Section 4.


\section{Brief review of massless DKP theory}


We start with the lagrangian density {for the} massless DKP fields in Minkowski%
\footnote{%
Throughout the text we adopt the convention that Latin indexes refer to the
Minkowski space-time with (constant) metric $\eta ^{ab}=(+1,-1,-1,-1)$,
while Greek indexes refer to  {a} riemannian space-time with metric $%
g^{\mu\nu}(x) $.} space-time \cite{HarishC, CQG}
\begin{equation}
\mathcal{L_{M}}=i\overline{\psi}\gamma\beta^{a}\partial_{a}\psi-i
\partial_{a}\overline{\psi}\beta^{a}\gamma\psi-\overline{\psi} \gamma\psi,
\label{eq1}
\end{equation}
where $\overline{\psi}=\psi^{\dagger}\eta^{0}\;,\quad \eta^{0}=2\!
\left(\beta^{0}\right)^{2}\!-\!1$, and $\gamma$ and $\beta^{a}$ are matrices
satisfying the massless DKP algebra\footnote{%
We choose a representation in which ${\beta^{0}}^\dag =\beta^{0}$, ${%
\beta^{i}}^{\dag}=-\beta^{i} $ and $\gamma^{\dag}=\gamma $\thinspace.}
\begin{subequations}
\label{eq3}
\begin{gather}
\beta^{a}\beta^{b}\beta^{c}+\beta^{c}\beta^{b}\beta^{a}=
\beta^{a}\eta^{bc}+\beta^{c}\eta^{ba}, \\
\beta^{a}\gamma+\gamma\beta^{a}=\beta^{a},\ \gamma^{2}=\gamma.
\end{gather}
The resulting equation of motion is
\end{subequations}
\begin{equation}
i\beta^{a}\partial_{a}\psi-\gamma\psi=0.  \label{eq2}
\end{equation}

In fact, the equations (\ref{eq1})-(\ref{eq2}) describe a set of
\textit{four} free massless gauge fields \cite{HarishC,Cox}, among
them the massless Klein-Gordon-Fock and the Maxwell electromagnetic
fields. The remaining two fields are of topological nature and will
not be considered here. The theory presented so far is manifestly
covariant under Lorentz transformations.

The above equations can be generalized to riemannian space-times
through the formalism of \textit{tetrads} (also called
\textit{vierbeins}), together with the \textit{minimal coupling
procedure}. Here we shall simply quote the main results we
 {shall} need; for details we refer to \cite{ijmpa,CQG} and
references therein. In this formalism we consider a riemannian
space-time $\mathcal{R}$ with metric $g_{\mu\nu}$, whose point
coordinates are labeled $x^{\mu}$. To each point in $\mathcal{R}$ we
consider a tangent Minkowski space-time $\mathcal{M}$ with
(constant) metric $\eta_{ab}$, whose point coordinates are labeled
$x^{a}$. The DKP fields $\psi$ are \textit{Lorentz group}
representations in this Minkowski space. The projections into
$\mathcal{R}$ of all tensor quantities defined on $\mathcal{M}$ are
done \textit{via} the tetrad fields $e^{\mu}{}_{a}(x)$. These fields
and its inverses $E_{\mu}{}^{a}(x)$ satisfy the relations
\begin{subequations}
\begin{align}
g^{\mu\nu}(x) & =\eta^{ab}e^{\mu}{}_{a}(x)e^{\nu}{}_{b}(x), \\
g_{\mu\nu}(x) & =\eta_{ab}E_{\mu}{}^{a}(x)E_{\nu}{}^{b}(x), \\
E_{\mu}{}^{a}(x) & =g_{\mu\nu}(x)\eta^{ab}e^{\nu}{}_{b}(x),
\end{align}
and
\end{subequations}
\begin{equation*}
e=\det\left( E_{\mu}{}^{a}\right) =\sqrt{-g},
\end{equation*}
where $g=\mathrm{det}(g_{\mu\nu})$.

The resulting lagrangian density for massless DKP fields minimally coupled
to $\mathcal{R}$ is
\begin{equation}
\mathcal{L_{R}}=e\left( i\overline{\psi }\gamma \beta ^{a}e^{\mu
}{}_{a}\nabla _{\mu }\psi -ie^{\mu }{}_{a}\nabla _{\mu }\overline{\psi }%
\beta ^{a}\gamma \psi -\overline{\psi }\gamma \psi \right) ,  \label{eq4}
\end{equation}%
where $\nabla _{\mu }$ is the riemannian covariant derivative
associated to the symmetric connection (Christoffel symbol) $\Gamma
_{\mu \nu }{}^{\alpha } $. The covariant derivatives of DKP fields
are
\begin{subequations}
\label{eq5}
\begin{align}
\nabla _{\mu }\psi & =\partial _{\mu }\psi +\frac{1}{2}\omega _{\mu
ab}S^{ab}\psi , \\
\nabla _{\mu }\overline{\psi }& =\partial _{\mu }\overline{\psi }-\frac{1}{2}%
\omega _{\mu ab}\overline{\psi }S^{ab},
\end{align}%
where $S^{ab}=[\beta ^{a},\beta ^{b}]$ and $\omega _{\mu ab}$ is the spin
connection, which in riemannian space-times is related to the metric and
vierbeins in the following way \cite{Bertlmann}
\end{subequations}
\begin{equation}
\omega _{\mu }{}^{ab}=e^{\nu b}\left\{ \frac{1}{2}E_{\rho }{}^{a}g^{\rho
\lambda }\left( \partial _{\mu }g_{\lambda \nu }+\partial _{\nu }g_{\lambda
\mu }-\partial _{\lambda }g_{\mu \nu }\right) -\partial _{\mu }E_{\nu
}{}^{a}\right\} \,.  \label{sc}
\end{equation}%
The equation of motion for massless DKP fields in $\mathcal{R}$ follows from
lagrangian (\ref{eq4})
\begin{equation}
i\beta ^{a}e^{\mu }{}_{a}\nabla _{\mu }\psi -\gamma \psi =0.  \label{eq6}
\end{equation}

\begin{sloppypar}
We recall that (\ref{eq4}) and (\ref{eq6}) are manifestly covariant
under \textit{Lorentz transformations} on {the \textit{tangent}
Minkowski space-time and also under \textit{general coordinates
transformations} on the underlying riemannian space-time}
\cite{ref4}.
\end{sloppypar}

\subsection{Spin 0 sector}


The scalar sector of massless DKP theory can be explicitly worked out by
using a five dimensional representation (see \cite{CQG}) of DKP algebra (\ref%
{eq3}). In this case the field $\psi$ is given by a 5-component column
vector
\begin{equation}  \label{eq22}
\psi=\left(\varphi,\psi^{0},\psi^{1},\psi^{2},\psi^{3}\right)^{T},
\end{equation}
where $\varphi$ and $\psi^{a}$ ($a=0,1,2,3$) behave respectively as a scalar
and a 4-vector under \textit{Lorentz} transformations on the Minkowski
tangent space. Expressing lagrangian (\ref{eq4}) in this representation we
obtain
\begin{equation}
\mathcal{L}_{0}=e\left(\frac{}{}\!i\psi^{\ast a} e^{\mu}{}_{a} \nabla_{\mu}
\varphi - i\psi^{a}e^{\mu}{}_{a}\nabla_{\mu} \varphi^{\ast}- \psi^{\ast
a}\psi_{a}\right) .  \label{eq24}
\end{equation}
The massless DKP equation (\ref{eq6}) implies
\begin{subequations}
\begin{gather}
\psi_{a}=ie^{\mu}{}_{a}\nabla_{\mu}\varphi,  \label{eq26a} \\[0.2cm]
e^{\mu}{}_{a}\nabla_{\mu}\psi^{a}=0 \, ,  \label{eq26b}
\end{gather}
which, together, give
\end{subequations}
\begin{equation}
g^{\mu\nu}\nabla_{\mu}\nabla_{\nu}\varphi=0,  \label{mkg}
\end{equation}
which is the massless Klein-Gordon-Fock equation in $\mathcal{R}$.
Accordingly, after substituting equation (\ref{eq26a}) into
lagrangian (\ref{eq24}), we obtain the corresponding massless scalar
Klein-Gordon-Fock lagrangian minimally coupled to $\mathcal{R}$
\begin{equation}
\mathcal{L}_{0}=\sqrt{-g}g^{\mu\nu}\nabla_{\mu}\varphi^{\ast}
\nabla_{\nu}\varphi.  \label{eq27}
\end{equation}


\subsection{Spin 1 sector}


Similarly, we can work out explicitly the vector sector of the theory by
using a representation (see \cite{CQG}) for the DKP algebra (\ref{eq3}). The
field $\psi$ is now a 10-component column vector
\begin{equation}  \label{eq28}
\psi=\left(\psi^{0},\psi^{1},\psi^{2},\psi^{3},\psi^{23},
\psi^{31},\psi^{12}, \psi^{10},\psi^{20},\psi^{30}\right)^{T},
\end{equation}
where $\psi^{a}$ ($a=0,1,2,3$) and $\psi^{ab}$ behave, respectively, as a
4-vector and an antisymmetric tensor under \textit{Lorentz} transformations
on the Minkowski tangent space. In terms of these components the lagrangian (%
\ref{eq4}) is written as
\begin{equation}
\mathcal{L}_{1}=e\left( \frac{i}{2}\psi^{\ast ab}\left(
e^{\mu}{}_{a}\nabla_{\mu}\psi_{b}-e^{\mu}{}_{b}\nabla_{\mu} \psi_{a}\right) -%
\frac{i}{2}\psi^{ab}\left( e^{\mu}{}_{a}
\nabla_{\mu}\psi_{b}^{\ast}-e^{\mu}{}_{b}\nabla_{\mu} \psi_{a}^{\ast}\right)
+\frac{1}{2}\psi^{\ast ab}\psi_{ab}\right),  \label{eq30}
\end{equation}
while the equations of motion are
\begin{subequations}
\begin{gather}
\psi_{ab}=-i\left( e^{\mu}{}_{a}\nabla_{\mu} \psi_{b}- e^{\mu}{}_{b}\nabla
_{\mu}\psi_{a}\right) ,  \label{eq32a} \\[0.2cm]
e^{\mu}{}_{b}\nabla_{\mu}\psi^{ab}=0.  \label{eq32b}
\end{gather}
Together, they give
\end{subequations}
\begin{equation}
\nabla_{\nu}F^{\mu\nu}=0,
\end{equation}
where
\begin{equation}
F_{\mu\nu}=\nabla_{\mu}\psi_{\nu}-\nabla_{\nu}\psi_{\mu}=
\partial_{\mu}\psi_{\nu}-\partial_{\nu}\psi_{\mu}.  \label{eq34}
\end{equation}
Correspondingly, turning the equation of motion (\ref{eq32a}) into the
lagrangian (\ref{eq30}), we obtain the Maxwell lagrangian minimally coupled
to $\mathcal{R}$\footnote{%
We identify $\psi_{\mu}$ with the electromagnetic field $A_{\mu}$.}
\begin{equation}
\mathcal{L}_{1}=-\frac{\sqrt{-g}}{2}g^{\mu\alpha}g^{\nu\beta}
F_{\mu\nu}^{\ast}F_{\alpha\beta}.  \label{eq33}
\end{equation}



\section{Conformal invariance}


In order to study the properties of massless DKP theory under conformal
transformations we construct the action corresponding to the lagrangian (\ref%
{eq4}),
\begin{equation}  \label{eq12}
S_{_{\mathcal{R}}}=\int d^{4}x\;e\left[ i\overline{\psi}\gamma
e^{\mu}{}_{a}\beta^{a}\nabla_{\mu}\psi-i\nabla_{\mu}\overline{\psi}
\beta^{a}e^{\mu}{}_{a}\gamma\psi-\overline{\psi}\gamma\psi\right] .
\end{equation}
As it had already mentioned, this action is manifestly invariant
under general coordinate transformations on the riemannian
space-time. Therefore, the study of conformal invariance of the
theory can be reduced to the study of the properties of lagrangian
(\ref{eq4}) under the so-called \textit{local Weyl rescalings}
\cite{buchbinder,Bertlmann}. In this case, invariance under local
Weyl rescalings imply conformal invariance.

In the following, as conventional in the study of local transformations \cite%
{blagojevic}, we first consider the restricted class of \textit{global} Weyl
rescalings in order to get the transformation laws for DKP fields. These
laws will be after generalized to the case of \textit{local} transformations.


\subsection{Global Weyl rescaling}


The restricted class of global Weyl rescalings of the metric and the \textit{%
vierbeins} is given by \cite{blagojevic, Bertlmann}
\begin{eqnarray}  \label{t1}
e^{\mu}{}_{a} & \rightarrow&{\mathrm{e}}^{-\sigma}e^{\mu}{}_{a} ,  \notag \\
E_{\mu}{}^{a} & \rightarrow&{\mathrm{e}}^{ \sigma}E_\mu{}^{a}, \\
e & \rightarrow& {\mathrm{e}}^{4\sigma}e,  \notag \\
g_{\mu\nu}&\rightarrow &{\mathrm{e}}^{2\sigma}g_{\mu\nu},  \notag
\end{eqnarray}
where $\sigma$, the \textit{scale parameter}, is an arbitrary real number.

In order to determine the transformation laws for the DKP field, we first
note that $\psi $ is represented by a column matrix \cite{dkp} whose
components have different canonical dimensions and, accordingly, must have
different conformal weights. To deal with this fact, we search for a
nonsingular real \textit{matrix} $M(\sigma )$, depending on the scale
parameter $\sigma $, such that under (\ref{t1}) $\psi $ behaves as
\begin{equation}
\psi \;\;\rightarrow \;\;M(\sigma )\psi ,  \label{eq14}
\end{equation}%
and, accordingly,
\begin{equation}
\overline{\psi }\;\;\rightarrow \;\;\overline{\psi }\eta ^{0}M^{T}(\sigma
)\eta ^{0}\;=\;\overline{\psi }M(\sigma ).  \label{eq14-1}
\end{equation}%
The \textquotedblleft contracted" spin conection
\begin{equation}
\omega _{\mu }\equiv \frac{1}{2}\;\omega _{\mu ab}\,S^{ab}  \label{wu1}
\end{equation}%
remains invariant under (\ref{t1}), as {it} can be shown from
(\ref{sc}). Accordingly, DKP covariant derivatives transform
covariantly, i.e.,
\begin{subequations}
\begin{align}
\nabla _{\mu }\psi & \;\;\rightarrow \;\;M(\sigma )\nabla _{\mu }\psi ,
\label{covder} \\[0.1cm]
\nabla _{\mu }\overline{\psi }& \;\;\rightarrow \;\;(\nabla _{\mu }\overline{%
\psi })M(\sigma )\;.
\end{align}%
The action (\ref{eq12}) will be invariant under the above transformations
if, and only if, the matrix $M(\sigma )$ satisfy the following conditions:
\end{subequations}
\begin{subequations}
\label{co12-0}
\begin{align}
e^{3\sigma }M(\sigma )\gamma \beta ^{a}M(\sigma )& \;=\;\gamma \beta ^{a}\;,
\\[0.1cm]
e^{3\sigma }M(\sigma )\beta ^{a}\gamma M(\sigma )& \;=\;\beta ^{a}\gamma \;,
\\[0.1cm]
e^{4\sigma }M(\sigma )\gamma M(\sigma )& \;=\;\gamma \;,
\end{align}%
which, with the aid of algebra (\ref{eq3}), can be reduced to
\end{subequations}
\begin{subequations}
\label{co12}
\begin{align}
M(\sigma )\beta ^{a}M(\sigma )& \;=\;e^{-3\sigma }\beta ^{a}\;, \\[0.1cm]
M(\sigma )\gamma M(\sigma )& \;=\;e^{-4\sigma }\gamma \;.
\end{align}%
It can be verified that the matrix
\end{subequations}
\begin{equation}
M(\sigma )=(1-\gamma )\mathrm{e}^{-\sigma }+\gamma \mathrm{e}^{-2\sigma }
\label{mm1}
\end{equation}%
satisfy all the above requirements and it is the solution we search for. For
instance, this matrix can be easily determined from the above conditions by
using the explicit representations for DKP algebra given in \cite{CQG}.

Summarizing, the massless DKP theory minimally coupled to riemannian
space-times is invariant under \textit{global} Weyl rescalings. Of
course, this holds for both spin 0 and spin 1 sectors of the theory,
as {it} can be easily verified.


\subsection{Local Weyl rescaling}


The \textit{local} Weyl rescalings can be obtained from the global
transformations (\ref{t1}), (\ref{eq14}) and (\ref{mm1}) by allowing the
scale parameter $\sigma$ to become a smooth function (the \textit{scale
function}) depending on the coordinates, denoted as $\sigma(x)$. It can be
easily verified that conditions (\ref{co12-0}) still hold in the local case.

From (\ref{sc}) and (\ref{wu1}) it can be shown that the spin connection $%
\omega _{\mu }$ now changes according to
\begin{equation}
\omega _{\mu }\rightarrow \omega _{\mu }+e^{\nu \,b}E_{\mu
}{}^{a}S_{ab}\,\partial _{\nu }\sigma \;,  \label{spin2L}
\end{equation}%
which now implies that {the} covariant derivatives of DKP field do
not, in general, transform covariantly. Instead, they transform as
\begin{equation}
\nabla _{\mu }\psi \rightarrow M\nabla _{\mu }\psi +\left( \partial _{\mu
}M\right) \psi +M\left( \partial _{\nu }\sigma \right) e^{\nu \,b}E_{\mu
}{}^{a}\,\,S_{ab}\psi .  \label{cdl}
\end{equation}%
Taking into account the explicit form (\ref{mm1}) for the matrix $M\!\left(
\sigma (x)\right) $ and the fact that $[M,S_{ab}]\!=\!0$, we find that the
first term in lagrangian (\ref{eq4}) changes as
\begin{equation}
i\,e\,\overline{\psi }\gamma \,\beta ^{\mu }\nabla _{\mu }\psi \rightarrow
i\,e\,\overline{\psi }\gamma \,\beta ^{\mu }\nabla _{\mu }\psi +i\,e\,%
\overline{\psi }\,\left( \frac{{}}{{}}\!\!-2\gamma +\gamma \beta _{a}\beta
^{a}\,\right) \,\beta ^{b}\,e^{\mu }{}_{b}\,\partial _{\mu }\sigma \,\psi \,,
\end{equation}%
where we have used the algebraic identity $\;\beta _{a}\beta ^{b}\beta
^{a}\!=\!\beta ^{b}\;$. Similarly, using the identity $\;\beta _{a}\beta
^{a}\beta ^{b}\!+\!\beta ^{b}\beta ^{a}\beta _{a}\!=\!5\beta ^{b}\;$, the
second term in the lagrangian changes as
\begin{equation}
i\,e\,\nabla _{\mu }\overline{\psi }\beta ^{\mu }\gamma \psi \rightarrow
i\,e\,\nabla _{\mu }\overline{\psi }\beta ^{\mu }\gamma \psi +i\,e\,%
\overline{\psi }\,\left( \frac{{}}{{}}\!\!-3\gamma +\gamma \beta _{a}\beta
^{a}-\beta _{a}\beta ^{a}+3\,\right) \,\beta ^{b}\,e^{\mu }{}_{b}\,\partial
_{\mu }\sigma \,\psi \,,
\end{equation}%
while the third term remains invariant.

Collecting the above results we find that the local Weyl rescalings change
the lagrangian (\ref{eq4}) by an amount $\delta \mathcal{L}$, given by
\begin{equation}  \label{conf-31-2}
\delta \mathcal{L}=i\,e\,\overline{\psi} \left(\frac{}{}\!\! \beta_a
\beta^a-3+ \gamma\right) e^\mu{}_b \beta^b \,\psi\, \partial_\mu\sigma\, .
\end{equation}
This variation can not be identically vanishing, because it would imply the
condition
\begin{equation}  \label{van}
\beta_a \beta^a\!-\!3\!+\!\gamma\!=0\, ,
\end{equation}
which is not an \textit{algebraic identity}. More precisely, though this
condition is satisfied restricted to the spin 1 sector of the theory, it
\textit{is not satisfied} in the spin 0 sector, where it is easy to verify%
\footnote{%
by using an explicit representation for the spin 0 sector, for example} that
holds the following relation
\begin{equation}  \label{zero}
\beta_a \beta^a-3+\gamma=1-2\gamma\, .
\end{equation}

Summarizing, we have shown that \textit{only the spin 1 sector} of the
massless DKP theory minimally coupled to riemannian space-times is
conformally invariant. Restricted to this sector, the transformation law (%
\ref{spin2L}) can be written as
\begin{equation}
\omega _{\mu }\rightarrow M\omega _{\mu }M^{-1}+M\partial _{\mu
}M^{-1}\;=\;\omega _{\mu }+(1+\gamma )\partial _{\mu }\sigma \,,
\label{conf-35}
\end{equation}%
and, accordingly, transformation (\ref{cdl}) implies that the covariant
derivatives transform covariantly,
\begin{eqnarray}
\nabla _{\mu }\psi  &\rightarrow &M(\sigma (x))\nabla _{\mu }\psi   \notag \\
\nabla _{\mu }\overline{\psi } &\rightarrow &(\nabla _{\mu }\overline{\psi }%
)M(\sigma (x))\,.
\end{eqnarray}

The above results are in complete agreement with those obtained in
the framework of (massless) KGF and Maxwell theories for the spin 0
and spin 1 sectors, respectively. This could be expected from the
equivalence between these theories and DKP when a minimal coupling
to riemannian space-times is considered \cite{CQG}. In the next
subsection, guided by the form of variation (\ref{conf-31-2}), we
achieve in a natural way the conformal invariance of also the spin 0
sector, without changing the content of the spin 1 sector.


\subsection{Conformal invariant theory}


We now require the whole theory to be invariant under local Weyl rescalings.
This can be achieved, as usual, by adding to the lagrangian (\ref{eq4}) a
compensating term, such that its variation cancels the variation (\ref%
{conf-31-2}) \cite{blagojevic}. The most natural form for such a term is
\begin{equation}
-i\,e\,\overline{\psi }\left( \frac{{}}{{}}\!\!\beta _{a}\beta
^{a}-3+\gamma \right) e^{\mu }{}_{b}\beta ^{b}C_{\mu }\,\psi \,\,,
\label{ct-44}
\end{equation}%
where $C_{\mu }$ must be a compensating field transforming under local Weyl
rescalings as
\begin{equation}
C_{\mu }\rightarrow C_{\mu }+\partial _{\mu }\sigma \,.  \label{ct-45}
\end{equation}

As we do not intend to introduce any new kind of field into the theory, we
require that $C_{\mu }$ must be given solely in terms of the metric tensor%
\cite{fulton}. Because (\ref{ct-44}) vanishes identically in the spin 1
sector, from now on we shall be concerned only with the changes it causes in
the spin 0 sector. Restricted to this sector, and taking into account
relation (\ref{zero}), the equation of motion reads
\begin{equation}
i\beta ^{\mu }\nabla _{\mu }\psi -\gamma \psi -i\,\left( 1-2\gamma \right)
\beta ^{\mu }C_{\mu }\,\psi =0\,.  \label{ct-48}
\end{equation}%
By using the scalar representation of reference \cite{CQG} this equation
gives the following relations among DKP field components:
\begin{subequations}
\label{ct-49}
\begin{align}
\left( \nabla _{\mu }-C_{\mu }\right) \psi ^{\mu }& =0\,, \\[0.2cm]
\psi _{\mu }& =i\left( \nabla _{\mu }+C_{\mu }\right) \varphi \,,
\end{align}%
which, together, result in the following equation for the scalar field $%
\varphi $
\end{subequations}
\begin{equation}
g^{\mu \nu }\nabla _{\mu }\partial _{\nu }\varphi -C\varphi =0\,,
\label{ct-50}
\end{equation}%
where the scalar $C$ was defined as
\begin{equation}
C\equiv g^{\mu \nu }\left( C_{\mu }C_{\nu }-\nabla _{\mu }C_{\nu }\right) \,.
\label{ct-51}
\end{equation}

Naturally, as $C_{\mu }$ is given solely in terms of the metric tensor, the
same holds for the scalar $C$. As a consequence, this scalar must be
completely determined as a function of the scalar curvature $R$ \cite%
{weinberg}. To determine precisely this function we observe that, under Weyl
rescalings
\begin{equation}
C\rightarrow e^{-2\sigma }\left( C-g^{\mu \nu }\nabla _{\mu }\partial _{\nu
}\sigma -g^{\mu \nu }\partial _{\mu }\sigma \partial _{\nu }\sigma \right)
\,,  \label{ct-52}
\end{equation}%
which is precisely the same way that the scalar $\frac{1}{6}R$ behaves under
these transformations. This result implies that
\begin{equation}
C=\frac{R}{6}\;.  \label{ct-54}
\end{equation}%
When this solution is replaced into (\ref{ct-50}), it gives the well-known $%
\frac{R}{6}\varphi $ coupling among scalar curvature and the
massless scalar KGF field.

\subsection{Effects of torsion}

{We now go a step further from riemannian to Riemann-Cartan
space-times by} discussing the effects of a nonvanishing torsion on
the scalar sector. We shall consider the ``weak" and ``strong" ways
{of defining} the conformal transformations of the torsion (see
\cite{Shapiro1} for a review, for instance). In the weak case the
torsion $Q^{\nu }{}_{\mu \lambda }$ (which is the antisymmetric part
of the spacetime connection) does not change under Weyl rescalings,
i.e.
\begin{equation}
Q^{\nu }{}_{\mu \lambda }\rightarrow Q^{\nu }{}_{\mu \lambda }\, .
\end{equation}
We recall that the spin connection in the presence of torsion is
given by \cite{CQG}
$$
\tilde{\omega}_{\mu }{}^{a}{}_{b} =\omega _{\mu
}{}^{a}{}_{b}-K^{a}{}_{b\mu }\, ,
$$
where $\omega _{\mu }{}^{a}{}_{b}$ is given by (\ref{sc}) and
$K^{a}{}_{b\mu }=E_{\nu}{}^a e^{\alpha}{}_b K^{\nu }{}_{\alpha \mu
}$, while
$$
K^{\nu }{}_{\alpha \mu }=Q^{\nu }{}_{\alpha \mu }-Q_{\alpha
}{}^{\nu }{}_{\mu }-Q_{\mu }{}^{\nu }{}_{\alpha }
$$
is the contorsion tensor. Accordingly, in the weak case the spin
connection changes as before, as if {there were} not torsion, namely
\begin{equation}
\tilde{\omega}_{\mu }\rightarrow \tilde{\omega}_{\mu }+e^{\nu
b}E_{\mu }{}^{a}S_{ab}\partial _{\nu }\sigma\, .
\end{equation}%
Therefore, all the  analysis previously done for the riemannian
space-times still holds in the weak case.

On the other hand, in the strong case the torsion transforms as
\begin{equation}
Q^{\nu }{}_{\mu \lambda }\rightarrow Q^{\nu }{}_{\mu \lambda
}+r_c\left( \delta _{\lambda }^{\nu }\partial _{\mu }\sigma -\delta
_{\mu }^{\nu }\partial _{\lambda }\sigma \right)\, ,
\end{equation}
where $r_c$ is an arbitrary parameter. {In this case the spin
connection transforms as}
\begin{equation}
\tilde{\omega}_{\mu }\rightarrow \tilde{\omega}_{\mu }+\left(
1-2r_c\right) e^{\nu \,b}E_{\mu }{}^{a}S_{ab}\,\partial _{\nu
}\sigma \, ,
\end{equation}
{where now appear an extra term depending on the parameter $r_c$},
\begin{equation}
\delta \mathcal{L}=i\,e\,\overline{\psi }\left[
\frac{{}}{{}}\!\!\beta
_{a}\beta ^{a}-3+\gamma +2r_c\left( 4-3\gamma -\beta _{a}\beta ^{a}\right) %
\right] e^{\mu }{}_{b}\beta ^{b}\,\psi \,\partial _{\mu }\sigma \,.
\end{equation}
In order to assure the conformal invariance of the theory we must
add to the lagrangian the following compensating term
\begin{equation}
\mathcal{L}_{ct}=-i\,e\,\overline{\psi }\left[
\frac{{}}{{}}\!\!\beta
_{a}\beta ^{a}-3+\gamma +2r_c\left( 4-3\gamma -\beta _{a}\beta ^{a}\right) %
\right] \beta ^{b}\,e^{\mu }{}_{b}~\tilde{C}_{\mu }\psi \, ,
\end{equation}
where $\tilde C_{\mu }$ is a compensating field transforming under
local Weyl rescalings as
\begin{equation}
\tilde C_{\mu }\rightarrow \tilde C_{\mu }+\partial _{\mu }\sigma
\,. \label{ct-52x}
\end{equation}

{From} (\ref{zero}) we observe that the term depending on the
parameter $r_c$ within the brackets vanishes in the spin 0 sector.
Therefore the torsion does not affect our previous result on the
scalar sector and $\tilde C_\mu$ is the same field introduced in the
torsionless case. This result could {already} be expected from the
complete equivalence between massless Klein-Gordon-Fock and the
scalar sector of the massless DKP theory in Riemann-Cartan
space-time \cite{CQG}, together with the results of references
\cite{Helayel1,Shapiro1}. Summarizing, {both massless scalar DKP and
KGF fields does not couple to the torsion and both of them are
conformally invariant in Riemann-Cartan space-times with the same
conformal coupling $\frac{1}{6}R\varphi ^{2}$}, {exactly as it
occurs }in the riemannian case.



\section{Conclusions}

{In this paper we studied the behavior under conformal
transformations of massless DKP fields in riemannian space-times}.
We started from the massless theory in the riemannian space-times,
as obtained from the formalism of \textit{vierbein} fields and from
the standard form of the \textit{minimal coupling} procedure. Taking
advantage from the fact that this theory is already manifestly
covariant under general coordinate transformations, {we carried out
the analysis of conformal transformations by} investigating only its
invariance properties under the special group of local Weyl
transformations. In this case, invariance under {these last group
of} transformations imply the conformal invariance of the theory.

We showed that, while the spin 1 (vector) sector of  {this} theory
is invariant under conformal transformations, the spin 0 (scalar)
sector is not. These results were in complete agreement with those
obtained in the framework of (massless) Klein-Gordon-Fock and
Maxwell theories in the context of minimal couplings, as it was
expected in the grounds of our previous results in reference
\cite{CQG}, in which we demonstrated the complete equivalence
between such theories and DKP when minimal couplings with riemannian
space-times are considered. In order to achieve also the conformal
invariance of the scalar sector, without modifying the vector
sector, we introduced a compensating term into the lagrangian which,
by its turn, carried a compensating field into the theory. {By
imposing the condition that this field were }completely determined
from the metric tensor, we naturally obtained the well-known
conformal coupling $\frac{1}{6}R\varphi ^{2}$ among the massless
scalar field and {the scalar curvature}.

Finally, we extended our analysis beyond the riemannian case by
discussing the effects of torsion on the scalar sector. We obtained,
for this case, the same conformal coupling obtained in the
riemannian case. This result was expected on the grounds of
\cite{CQG} and in agreement with \cite{Helayel1,Shapiro1}.

\subsection*{Acknowledgements}

We are very grateful to {the anonymous referees}, whose critical
comments helped us to clarify several points {and to improve the
quality o}f this work. R.C. thanks FAPESP (grant 01/12611-7) for
full support; B.M.P. thanks CNPq and FAPESP (grant 02/00222-9) for
partial support; J.T.L. thanks MCT/CNPq/Funda\c{c}\~ao Arauc\'aria
(grant 1261) and also thanks A.C. Aguilar for her kind help in
obtaining several references.


\end{document}